\documentclass[journal,transmag]{IEEEtran}

\usepackage{makecell}
\usepackage{graphicx}
\usepackage{subcaption}
\usepackage{amsmath,amsfonts}
\usepackage{algorithmic}
\usepackage{algorithm}
\usepackage{array}
\usepackage{textcomp}
\usepackage{stfloats}    
\usepackage{url}
\usepackage{verbatim}
\usepackage{cite}
\usepackage{caption}
\usepackage{comment}
\usepackage{multirow}
\usepackage{booktabs}

\begin{document}

\title{Reinforcement Learning for Motor Control:\\ A Comprehensive Review}

\author{\IEEEauthorblockN{Danial Kazemikia}
\IEEEauthorblockA{Department of Electrical Engineering, University of Texas at Dallas, Richardson, TX 75080 USA}% 
\thanks{Corresponding author: D. Kazemikia (email: danialkazemikia@utdallas.edu).}}

\IEEEtitleabstractindextext{%
\begin{abstract}
Electric motors are crucial in many applications, but traditional control methods struggle with nonlinearities, parameter uncertainties, and external disturbances. Reinforcement Learning (RL) offers a promising solution as a data-driven approach that can learn optimal control strategies without an explicit model. This review paper examines the current state of RL in motor control, exploring various RL algorithms, and applications. The review highlights RL's advantages, including model-free control, adaptability to changing conditions, and the ability to optimize for complex objectives. It also addresses challenges in applying RL to motor control, such as sim-to-real transfer, safety and stability concerns, scalability, and computational complexity. By providing a comprehensive overview of the field, this review aims to deepen understanding of RL's potential to revolutionize motor control and drive advancements across industries.\end{abstract}

}

\maketitle

\IEEEdisplaynontitleabstractindextext

\IEEEpeerreviewmaketitle

\section{Introduction}
\IEEEPARstart{E}{lectric} motors are indispensable across a wide range of industries, including manufacturing, transportation, aerospace, and robotics. Precise control of electric motors is critical for ensuring high performance, energy efficiency, and system reliability. However, the complexity of motor control tasks continues to increase as systems demand greater precision, fault tolerance, and efficiency. Conventional control techniques—such as proportional-integral-derivative (PID) control and model predictive control (MPC) have traditionally formed the backbone of motor control strategies \cite{2,3,5,6,7,12}. While effective in many scenarios, these methods are inherently dependent on accurate mathematical models of the motor and its dynamics to predict system behavior and design appropriate control strategies, which can introduce significant limitations in real-world applications \cite{2, 6, 8, 10, 12, 13}.
Achieving an exact mathematical model is challenging in systems with complex dynamics, changing parameters, nonlinearities, or unmeasurable states \cite{2,3,5,6,7,12,13,14,15,16}. Thus, motor control systems often struggle with model uncertainty and parameter variability, especially in complex drives like permanent magnet synchronous motors (PMSMs) \cite{15}. The discrepancies, between a PMSM and its model, can degrade the performance and robustness of the control system,  especially due to the coupling effects between the dq-axis, nonlinearities, and measurement errors which hinder the effectiveness of conventional linear controllers such as PI controllers, resulting in suboptimal performance \cite{3}. 
Even when accurate models are available, parameter drift—arising from temperature fluctuations, aging, and mechanical wear can degrade the performance of model-based controllers over time \cite{8,9,10}. This is because many conventional controllers are designed for fixed operating conditions, making them less adaptable to changing environments or disturbances \cite{5}. Moreover, Advanced techniques such as MPC also introduce \textbf{computational challenges} due to real-time optimization requirements, which can be prohibitive for systems with limited computational resources \cite{14,16}

Given these challenges, there is increasing interest in \textbf{data-driven, model-free control strategies}, with \textbf{reinforcement learning (RL)} emerging as a promising solution. Unlike traditional control methods that rely on explicit models, RL enables the development of optimal control policies through interactions with the environment. In RL, an agent learns by observing the system’s states, taking actions, and receiving rewards based on performance outcomes \cite{1,2,5,6,10}. Over time, the RL algorithm identifies strategies that maximize cumulative rewards, achieving desired control objectives without requiring precise mathematical models \cite{19,7,16}. This relatively new method offers benefits over traditional control techniques, in motor control applications, which is going to briefly discuss hereunder: 

\begin{itemize}
    \item \textbf{Model-Free}: conventional control strategies like MPC or PI control rely on an accurate model of the motor, which can be difficult to obtain, particularly with real-world variations and complexities \cite{5,4}. However, RL eliminates the need for detailed system models~\cite{14}. This model-free characteristic makes RL particularly suitable for complex and uncertain environments where obtaining precise system models can be challenging or impractical \cite{10,7,6,5,4}.

    \item \textbf{Dynamic Adaptability}: RL agents benefit from the inherent ability to adjust to changing systems and environmental conditions \cite{1,4,6,12,14,15}. This adaptability ensures consistent performance even when facing alterations in parameters caused by temperature fluctuations, aging effects, unexpected disturbances (eg. variations in load torque and rotational inertia), etc. \cite{19}. Conventional controllers often demand thorough recalibration and parameter adjustments to sustain performance under such variations, making them less robust in practical settings.

    \item \textbf{Optimal Control}: RL's core principle involves maximizing a reward function, which encapsulates the desired control objectives. RL agents, through iterative learning processes, identify efficient control strategies precisely tailored to these specific, potentially complex performance metrics. This contrasts with conventional controllers that may struggle to accommodate intricate or unconventional performance criteria \cite{1,3,4}.

    \item \textbf{Handling Nonlinearities}: Another challenge in motor control stems from the intrinsic nonlinearities within electric motors and their associated power electronics. These nonlinearities frequently present difficulties for traditional control approaches, hindering their effectiveness. RL tackles these nonlinearities by directly learning complex, nonlinear relationships from input and output data. This data-driven nature empowers RL to effectively manage nonlinearities that are difficult for conventional controllers to address. \cite{10,16,17,18}
\end{itemize}

This review paper provides a comprehensive overview of the applications of RL in electric motor control. It explores various RL algorithms employed in this domain, categorizing them based on their learning mechanisms (value-based, policy-based, and actor-critic) and highlighting their strengths and limitations. The review discusses specific applications of RL in controlling different motor types, including DC motors, PMSMs, SRMs, and induction motors, emphasizing the achieved performance improvements, robustness enhancements, and efficiency gains. Finally, the paper analyzes the challenges and future research directions in the field of RL for motor control, focusing on issues such as sim-to-real transfer, safety and stability, scalability and computational complexity, and explainability and interpretability of RL-derived control policies.

\section{Fundamentals of Reinforcement Learning}

Reinforcement Learning (RL) is a machine learning approach where an agent learns optimal actions by interacting with an environment \cite{16,18}. Unlike supervised learning, which needs labeled datasets, RL happens through trial-and-error whereby the agent receives feedback in the form of rewards contingent upon its actions \cite{1}. This makes RL suitable for problems where defining explicit rules or providing comprehensive training data is difficult.

The foundation of RL is the concept of a \textbf{Markov Decision Process (MDP)}, which provides a mathematical structure for modeling decision-making problems. An MDP is characterized by the following elements:

\begin{itemize}
    \item \textbf{Agent:} The agent serves as the learner and decision-maker, observing the environment, executing actions, and receiving rewards based on its performance \cite{5,11,12}. 
    \item \textbf{Environment:} The environment represents the external system with which the agent interacts; it provides the agent with observations of its current state and responds to the agent's actions by transitioning to new states and issuing rewards \cite{18,1,4}.
    \item \textbf{States:} The state, $s$, characterizes the environment at a particular time, offering critical information about the current situation that the agent utilizes to make decisions \cite{1,6,17}.  In the context of motor control, a state might include the motor's speed, position, or current. \cite{13,14}
    \item \textbf{Actions:} Actions, $a$,  are the choices the agent can make to influence the environment. For instance, an agent might adjust the voltage applied to a motor \cite{2,6,8,13}, or switch commands for inverters \cite{7,15}. The primary objective of the agent is to select actions that yield favorable outcomes.
    \item \textbf{Policy:} The policy, $\pi$, represents a mapping from states to actions \cite{1,2,4,6,12}. It defines the agent's behavior, dictating the agent's decision-making strategy.  Policies can be either deterministic—yielding the same action for a given state—or stochastic, where actions are selected based on a probability distribution. 
    \item \textbf{Reward:} The reward, $r$, is a numerical feedback signal provided by the environment to the agent post-action, indicating the desirability of the action's outcome \cite{2,6,7,12}. Rewards may be positive, negative, or zero, driving the learning process and encouraging the agent to achieve the intended control objectives. For example, a high reward could be given for achieving a target speed, while exceeding current limits could lead to a penalty. As an example, \ref{reward} demonstrates a reward function, defined to minimize the speed error.

    \begin{figure}[!t]
    \centering
    \includegraphics[width=3.5in]{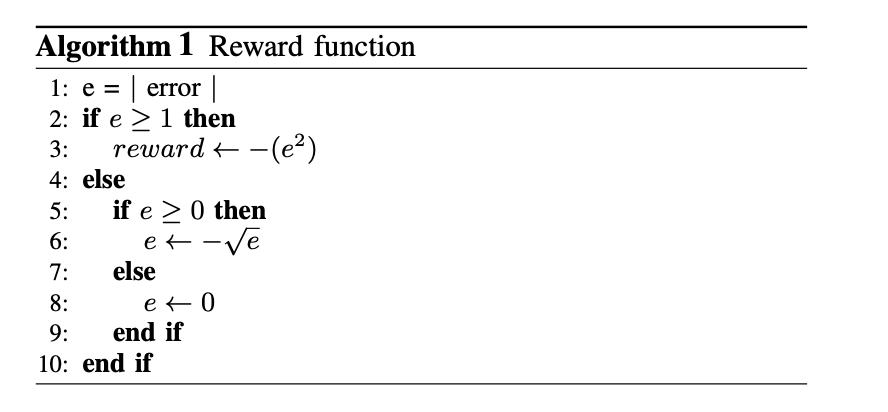}
    \caption{A framework of RL learning process. \cite{2}}
    \label{reward}
    \end{figure}

    \item \textbf{Discount Factor:} The discount factor, symbolized as $\gamma$ and ranging from 0 to 1, is a parameter used in reinforcement learning that determines the importance of future rewards relative to immediate rewards. A higher discount factor encourages the agent to prioritize long-term cumulative rewards, while a lower discount factor leads to a more myopic focus on immediate rewards \cite{1,7,12,9}. 

\end{itemize}

\begin{figure}[!t]
\centering
\includegraphics[width=3.5in]{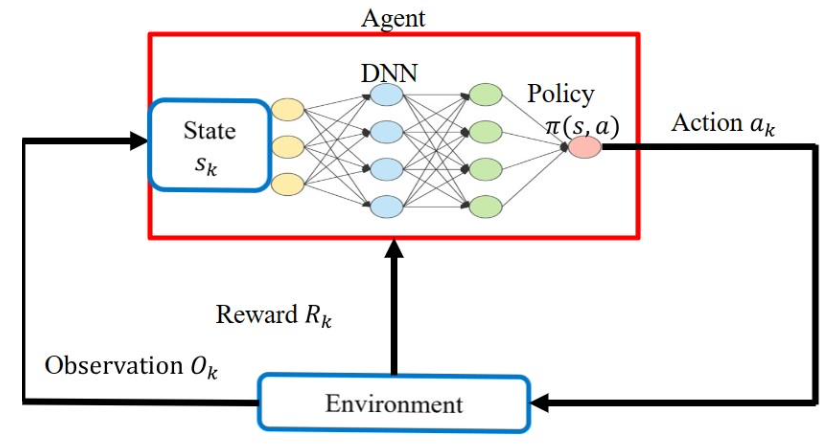}
\caption{A framework of RL learning process.
}
\label{RFMG}
\end{figure}

The goal in RL is for the agent to learn a policy that maximizes the expected cumulative reward over time. This learning process is iterative, involving the agent taking actions, observing the outcomes, receiving rewards, and then refining its policy accordingly \cite{1,2,3}.

The integration of deep learning has further transformed RL, enabling it to handle complex, high-dimensional state and action spaces. \textbf{Deep Reinforcement Learning (Deep RL)} employs deep neural networks to approximate value functions or policies, allowing RL agents to tackle challenging problems like game-playing, robotics, and autonomous systems with unprecedented success \cite{1,18}. This synergy between RL and deep learning has unlocked new possibilities, making RL a powerful tool for solving problems across a wide range of domains.

\section{Motor-Specific Applications of RL for Motor Control}

Previous studies have explored applications of RL to control different types of motors, which are briefly mentioned hereunder.

\subsection{Permanent Magnet Synchronous Motors (PMSMs)}
PMSMs are commonly used in various industrial applications due to their advantages such as high efficiency, power density, and compact structure. PMSM is commonly studied due to its simplicity and ease of control. However, they still present control challenges due to their nonlinear dynamics and susceptibility to internal parameter changes and external disturbances like load torque and rotational inertia variations. PMSMs are frequently studied in the context of RL-based control, with multiple sources highlighting the effectiveness of RL techniques in achieving precise control, especially for current regulation. \cite{1,3,4,6,7,8,9,12,14,15,18,19}

\subsection{Brushless DC Motors (BLDCMs)}
Some advantages of BLDCMs, such as fast response, easy adjustment, and stable performance contribute to their widespread use in various applications. \cite{16} demonstrates that integrating DDPG with a PID controller can significantly improve the speed-tracking accuracy of BLDCMs compared to using PID alone.

\subsection{DC Motors}
While PMSMs and BLDCMs dominate RL research, RL-based control is also applied to traditional DC motors, mainly for speed control \cite{2}.

\subsection{Switched Reluctance Motors (SRMs)}
SRMs are a type of electric motor known for their simple and robust structure, and they demonstrate high efficiency at high speeds. SRMs are gaining popularity in applications like electric vehicles and aerospace due to their resilience, high-temperature tolerance, and wide speed range. However, they have complex control requirements due to nonlinear characteristics \cite{10}.

\section{Types of RL Algorithms in Electric Motor Control}

A variety of RL algorithms are employed for diverse motor control applications. Categorizing based on the approach taken for learning optimal policies provides insights into their strengths and weaknesses and helps researchers select appropriate algorithms for specific tasks.

\subsection{Value-Based Methods}
These algorithms focus on learning a value function, which predicts the expected cumulative reward for each state or state-action pair. The objective is often to learn the optimal value function, which can subsequently be used to derive an optimal policy. These methods represent the policy implicitly by selecting actions that maximize the value function \cite{6}.

\textbf{Deep Q-Network (DQN):} As a Value-Based Method, DQN utilizes a deep neural network to approximate the Q-function, which estimates the long-term reward for given actions and observations. \cite{7,15,19} showcases the application of DQN in handling complex, high-dimensional state and action spaces in motor control problems.

\subsection{Policy-Based Methods}
In contrast to value-based methods, policy-based methods directly search for an optimal policy without explicitly modeling the value function \cite{6}. 

\textbf{Policy Gradient (PG):} Policy gradient methods, a well-known Policy-Based Method, update policy parameters by following the gradient of an objective function that assesses the policy's performance \cite{4,8,9,12,16,18}.

\subsection{Actor-Critic Methods}
Actor-critic methods combine value-based and policy gradient methods by utilizing two components: an actor-network that learns the policy, and a critic network that estimates the value function. The critic's role is to evaluate the actor's policy, offering feedback to enhance the policy's effectiveness [8, 9, 13, 14].

\textbf{Deep Deterministic Policy Gradient (DDPG):} DDPG extends DQN to continuous action spaces, making it well-suited for controlling motor variables like voltage and current, which are continuous signals. Several sources demonstrate the application of DDPG for motor control  \cite{4,8,9,12,16,18}.

\textbf{Twin Delayed DDPG (TD3):} TD3 is an improved variant of DDPG that addresses overestimation bias in the critic network, leading to more stable and reliable learning. Studies \cite{1,8,12,13}mention TD3 as a potential algorithm for motor control applications. In particular, \cite{12} compares the performance of DDPG and TD3 for disturbance rejection in a motion control system, suggesting that TD3 might offer better performance.

\section{Control Objectives in Motor Control using RL}

Reinforcement Learning (RL) has emerged as a promising technique in electric motor control, mainly utilized in two ways: direct motor control and the optimization of conventional controllers. A structured overview of existing research can be achieved by categorizing studies based on their control objectives, offering insight into the diverse capabilities of RL in handling various motor control challenges.

\subsection{Direct Motor Control Applications}

In this category, RL is used to develop a control policy that translates the motor's state into desired actions, such as adjusting the voltage applied to the armature of a DC motor to regulate speed \cite{2}. In various studies, RL has been demonstrated to be excellent at controlling parameters like speed, torque, and current directly in electric motors \cite{3,5,7,15}.

\cite{2} explores the application of a DQN algorithm for speed control of a DC motor by minimizing the error between the actual and reference speed. the DQN-based speed controller completely replaces a PID controller that provides the armature voltage signal to be sent to a DC motor. The results indicate that RL controllers can outperform PID controllers in terms of accuracy and response time for DC motor speed control, especially in dynamic environments with disturbances. However, the computational cost of training the DQN agent can be substantial and might limit its applicability in real-time systems.

\begin{figure}[!t]
\centering
\includegraphics[width=3in]{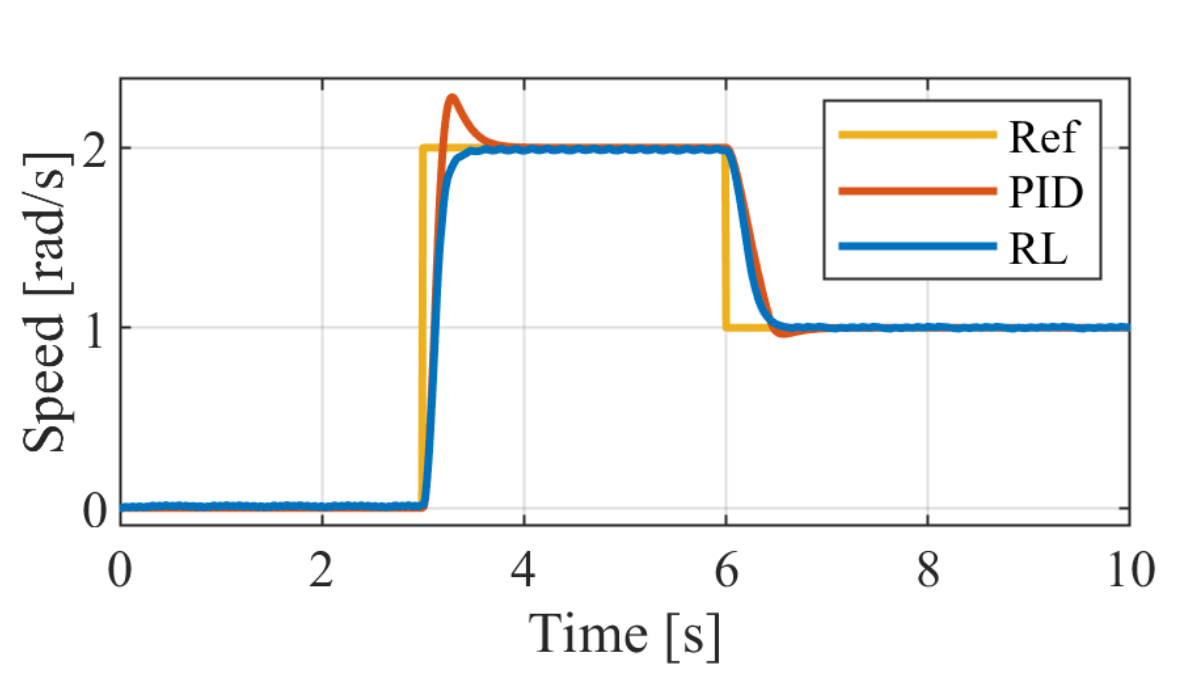}
\caption{Speed profile of PMSM controlled by PID vs RL \cite{19}}
\label{19.1}
\end{figure}

\begin{figure}[!t]
\centering
\includegraphics[width=3in]{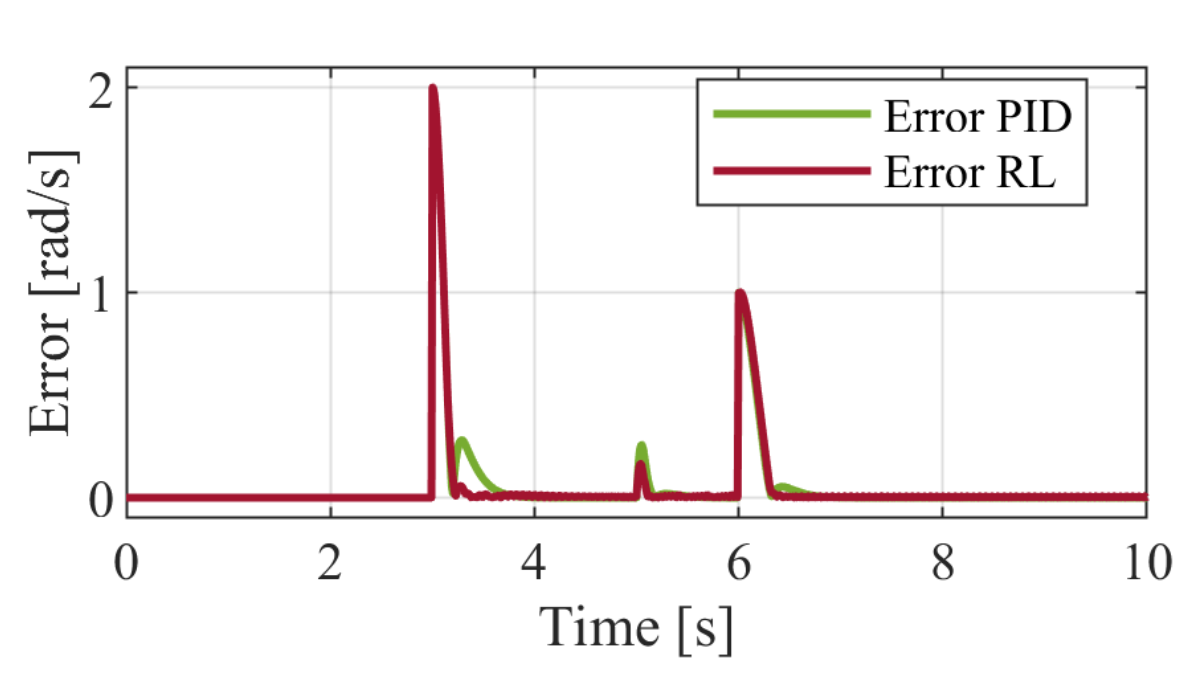}
\caption{Speed Error in PMSM controlled by PID vs RL \cite{19}}
\label{19.2}
\end{figure}

\cite{19} uses DQN to achieve accurate speed tracking of a PMSM, under load disturbances. The paper presents simulation results comparing the performance of the DRL controller with a traditional PI controller under different operating conditions, demonstrating the superior tracking performance and robustness of the DRL approach. \ref{19.1} and \ref{19.2} depicts the speed profile and the speed error in PMSM in this study.

\cite{3} This source aims to achieve robust current regulation in Surface-Mounted PMSM drives, ensuring accurate current control despite parameter uncertainties, and errors. It proposes an Integral Reinforcement Learning (IRL)-based control algorithm that uses real-time data for online learning to address the limitations of traditional offline RL methods, which are heavily dependent on pre-sampled data quality. The IRL algorithm uses an Actor-Critic neural network for direct control of the inner current loop. The outer speed loop uses a traditional PI controller. The developed control strategy is shown to surpass the performance of conventional Field-Oriented Control (FOC) with PI controllers, particularly in mitigating the cross-coupling effects between d-q axis currents.

\begin{figure}[!t]
\centering
\includegraphics[width=3.4in]{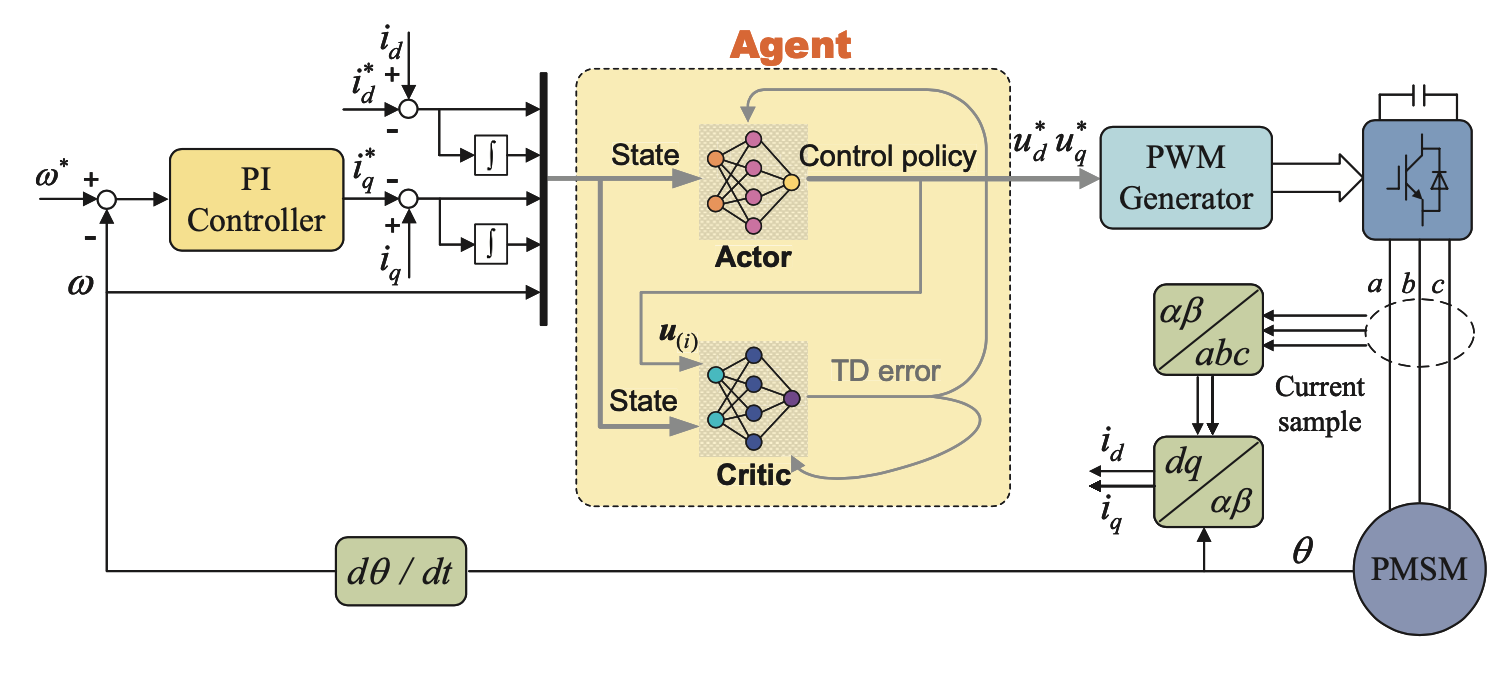}
\caption{Block Diagram of the PMSM robust control based on proposed online off-policy IRL Algorithm \cite{3}
}
\label{RFMG}
\end{figure}

\cite{6}  implements an actor-critic algorithm to regulate the d/q current components in a PMSM, which involves controlling the magnitude and phase of the motor currents to achieve the desired torque and speed in a field-oriented framework. The DRL-based controller outputs appropriate input voltages to reach the desired current values. The results however are limited to simulations.

\cite{7} Leverages a DQN-based method to minimize the stator current to maximize the efficiency while maintaining the commanded reference torque in a PMSM. This objective is formulated as a discrete-time dynamic optimization problem, aiming to find the optimal switching state for the inverter that minimizes stator current while adhering to operational constraints like current and voltage limits. On the other hand, the DQ-DTC does not consistently achieve zero d-axis current to achieve the highest possible steady-state efficiency. Moreover, using FPGA, as the embedded device, significantly adds to the hardware cost. 

\begin{figure}[!t]
\centering
\includegraphics[width=3.4in]{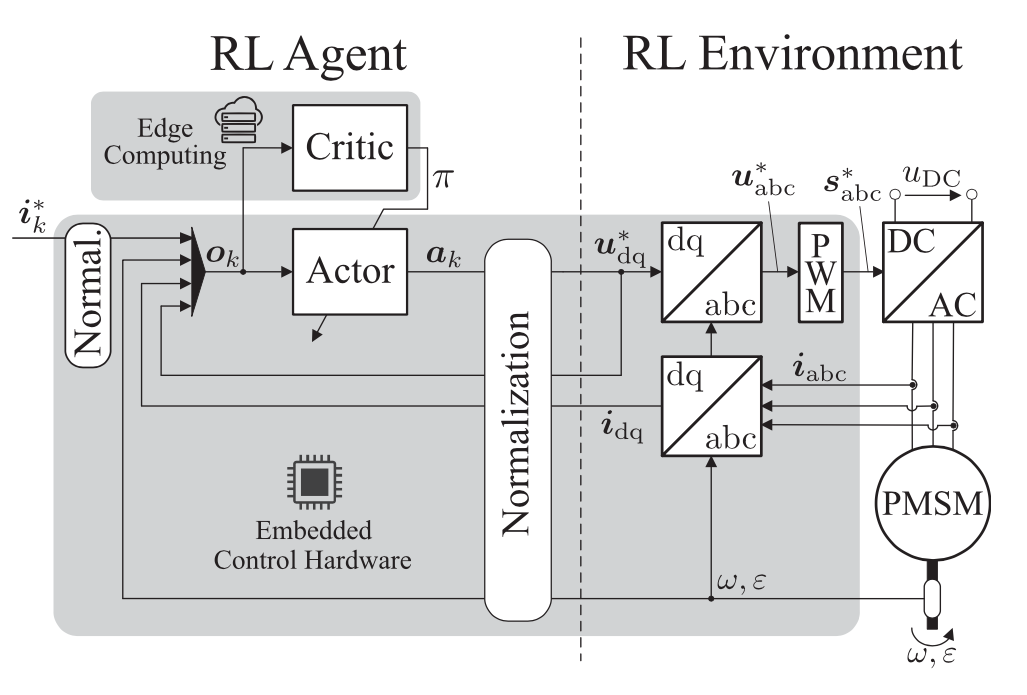}
\caption{Simplified schematic of the overall control and drive system in \cite{7}
}
\label{RFMG}
\end{figure}

\cite{10} The objective is to accurately track a reference current trajectory for a switched reluctance motor (SRM) drive. It proposes a Q-learning scheduling algorithm combined with a linear quadratic tracker (LQT) to handle the highly nonlinear electromagnetic nature of SRMs, which depends on variations in both phase current and rotor position. The Q-learning algorithm is used to select appropriate LQTs from a table based on the rotor angle and current, enabling adaptive and robust direct control of the SRM current. The controller regulates the pulse-type current characteristic of SRMs, demanding high current variations (di/dt) and a high bandwidth drive system. This study utilizes a table of Q-cores and a linear interpolation mechanism to achieve nonlinear tracking capabilities instead of using neural networks, which makes it computationally simpler than some neural-network-based RL methods. Moreover, although the algorithm adapts to variations from the expected model, it assumes that the underlying SRM parameters remain constant within each Q-core region.

\cite{8} introduces a TD3-based Meta-Reinforcement Learning (MRL) method for regulating the dq current components of a Permanent Magnet Synchronous Motor (PMSM) by controlling the voltage and frequency supplied to the motor through a two-level inverter with PWM. The authors propose a context-based off-policy MRL approach, inspired by PEARL and Meta-Q-learning, which includes a context network to process motor-specific observations and produce context variables that adapt the controller to different PMSM types without requiring separate training. However, the study notes several challenges: potential overfitting to commonly represented motor classes in the dataset, increased computational cost due to frequent context network updates, and limitations arising from the assumption of static motor parameters, which may not account for real-world variations like temperature changes and aging. The paper also suggests that the learned context variables might aid in fault detection, as deviations in these variables could indicate parameter changes due to faults.

\cite{15} presents a Multi-Set Robust Reinforcement Learning (MSR-RL) method, based on Deep Q-Networks (DQN), designed for current control in surface-mounted PMSM. MSR-RL offers robustness to parameter uncertainties and mismatches by training across diverse parameter sets, achieving a control policy that accurately tracks reference currents and ensures smooth operation under varied conditions. Representing the PMSM system as a Contextual Markov Decision Process (CMDP), this method clusters similar contexts into models, facilitating generalization to unseen parameters. Offline training is used, though this approach can be computationally intensive, and real-world performance may be affected by unmodeled dynamics. Validation through standard RL comparison demonstrates MSR-RL’s adaptability and robustness, though a direct comparison with a standalone PID controller could further elucidate its benefits.

\begin{figure}[!t]
\centering
\includegraphics[width=3.4in]{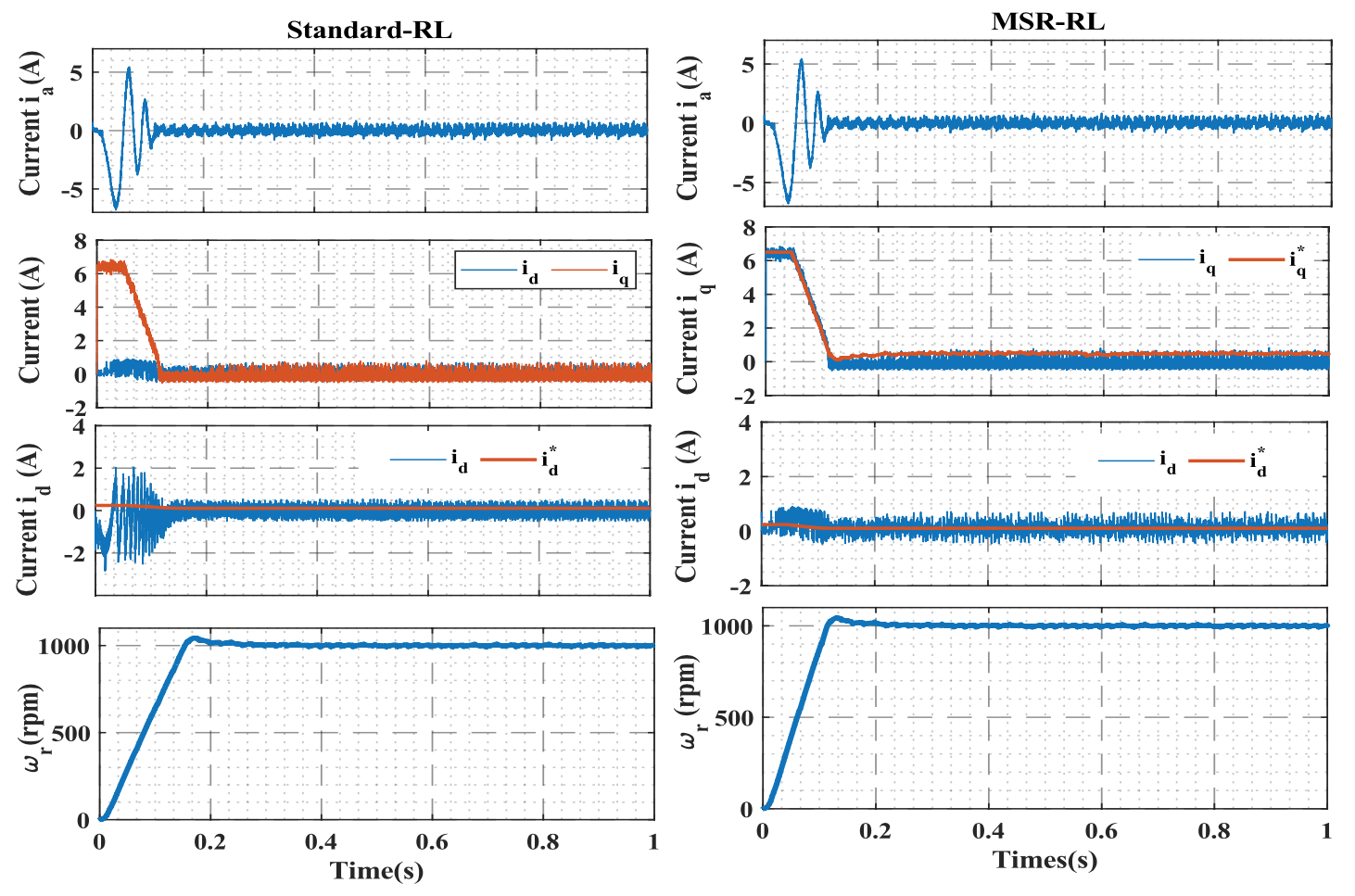}
\caption{Phase Current, q/d Currents, and Speed of a PMSM controlled by MSR-RL vs standard RL \cite{15} }
\label{RFMG}
\end{figure}

\cite{12} aims to minimize tracking errors caused by disturbances in motion control systems. DRL agents are trained to output a control signal combined with a classic closed-loop controller to improve the effectiveness of control.
The study employs two DRL algorithms, DDPG and TD3, to design data-driven controllers for disturbance rejection in motion control systems. A comparative analysis of the performance of DDPG and TD3 in disturbance rejection highlights the superiority of TD3 in continuous control tasks in motion control systems, over DDPG. The proposed algorithms are validated through simulations, and further experimental validation on real-world motion control systems is needed to assess their practical feasibility and limitations.

\begin{figure}[!t]
\centering
\includegraphics[width=3.4in]{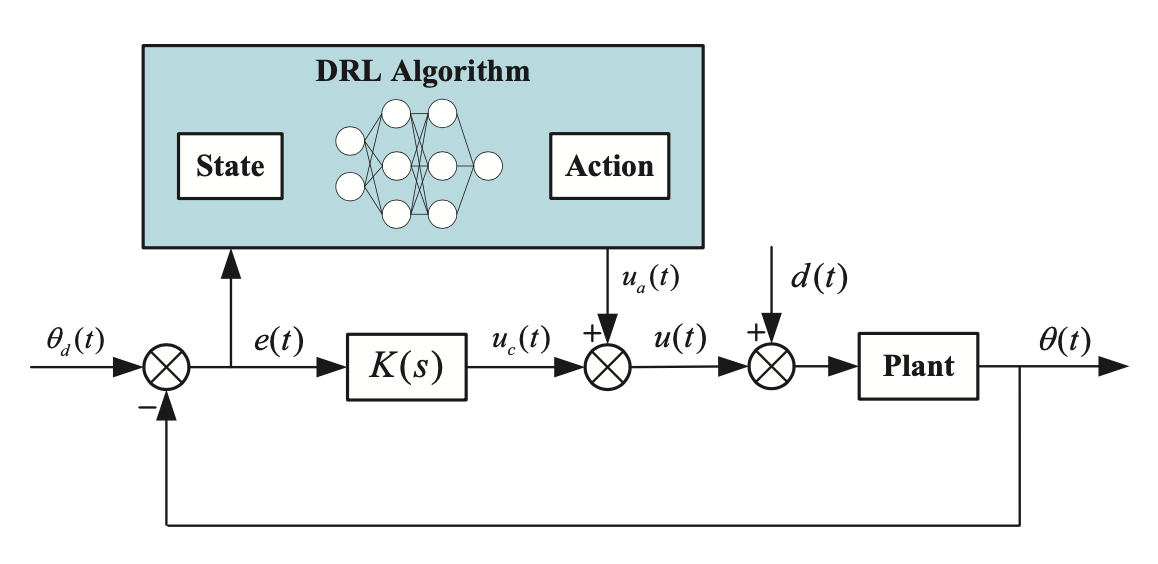}
\caption{Configuration of RL-based data-driven system \cite{12}.}
\label{RFMG}
\end{figure}

\begin{figure}[!t]
\centering
\includegraphics[width=3.4in]{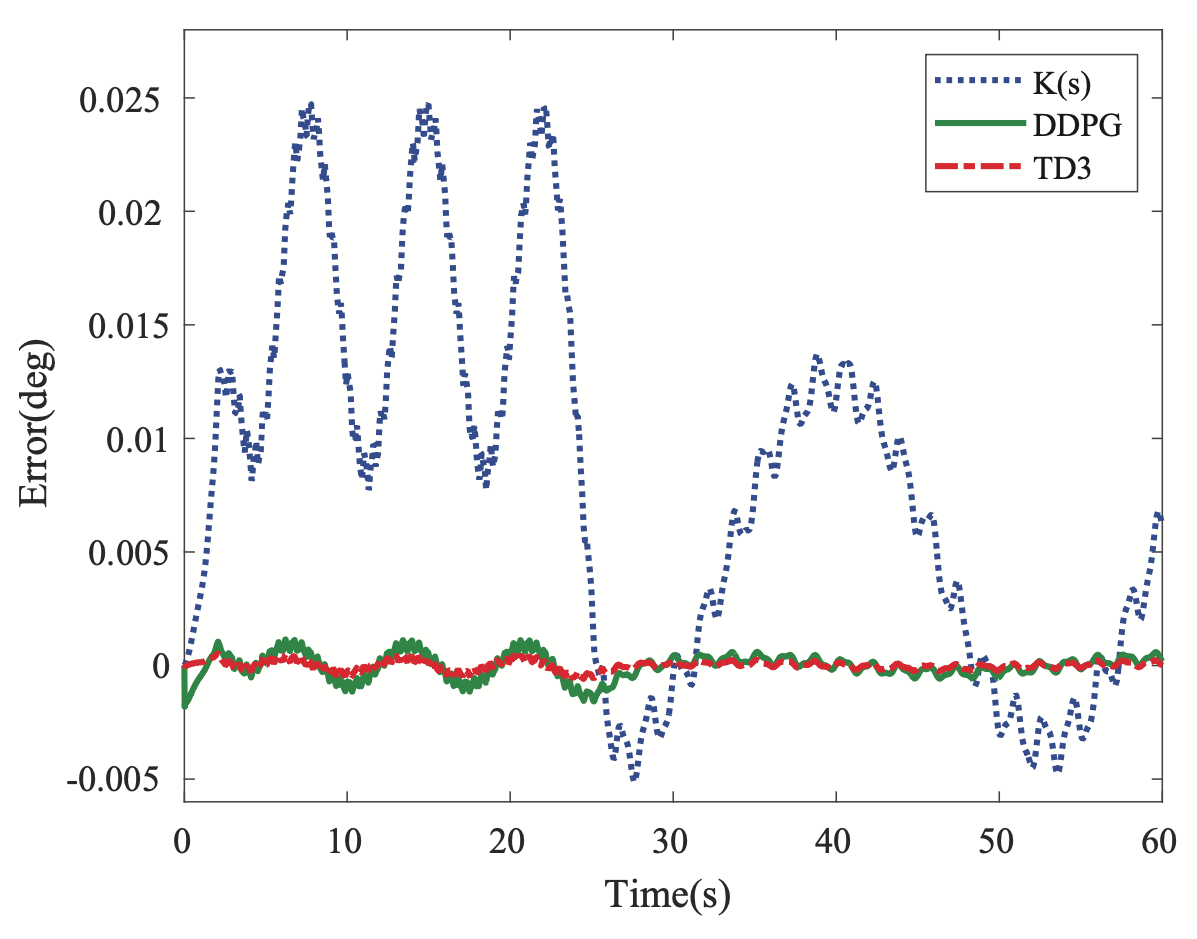}
\caption{Tracking error under K(s), DDPG and TD3 with disturbances \cite{12}.}
\label{RFMG}
\end{figure}

\subsection{Parameter Optimization}
In addition to direct control, RL significantly enhances the optimization of conventional motor controllers, such as PID controllers, by training an agent to identify optimal parameters based on the motor's state, thereby improving traditional system performance. This approach enables automatic tuning of controller parameters, which is essential for achieving desired performance and allows the system to adapt to varying operating conditions and uncertainties.

\cite{1} seeks to enhance the disturbance rejection ability and response speed of a PMSM motor used in a More Electric Aircraft (MEA). The TD3 algorithm is used to automatically optimize the parameters of the active disturbance rejection controller (ADRC), which then directly controls the motor, autonomously, bypassing the need for manual tuning. Simulations and experiments both show improved disturbance rejection capabilities and speed response of TD3-ADRC compared to traditional ADRC and Model Predictive Control (MPC).

\begin{figure}[!t]
\centering
\includegraphics[width=3.4in]{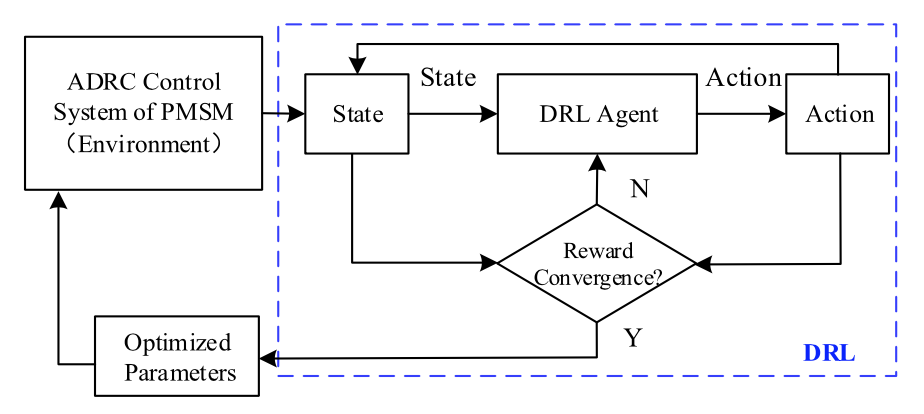}
\caption{Block diagram of parameter optimization using RL in \cite{4}.}
\label{4}
\end{figure}

\cite{4} aims to achieve robust speed control in the flux weakening control (FWC) system of an MEA PMSM, maintaining stable speed even at high speeds where magnetic flux weakening is applied. The authors focus on mitigating disturbances that can affect speed control. The DDPG algorithm is used to automatically tune the 11 parameters of the ADRC that control the motor. \ref{4} displays the block diagram of parameter optimization system in \cite{4}.

\cite{9} introduces the MSPO-DRL method, which uses DDPG to optimize the ADRC parameters for multiple control scenarios. This method achieves the desired speed-tracking performance for a PMSM, under various operating conditions, including different speeds, loads, and disturbances. MSPO-DRL also allows for accurately following a given speed trajectory and overcoming the limitations of traditional ADRC parameter tuning methods, such as trial-and-error and heuristic algorithms, which are time-consuming and struggle to adapt to different control scenarios. However, \cite{9} assumes that the PMSM parameters remain constant, neglecting potential variations caused by factors like temperature and aging.

\cite{14} aims to design a controller that can asymptotically regulate the speed of a PMSM to its setpoint, ensuring the motor speed gradually converges to and stays at the desired speed. The RL algorithm is used to learn the optimal controller parameters for speed control of the PMSM. The paper assumes a constant load torque. The authors used singular perturbation theory to decouple the fast and slow dynamics of the PMSM. This simplifies the outer loop controller design by reducing the model order. 

\cite{16} seeks to improve the speed tracking accuracy of a BLDCM, ensuring the motor's speed closely follows a predefined trajectory. It proposes combining a traditional PID controller with a DDPG algorithm. The DDPG algorithm monitors the BLDCM's state and adjusts the PID controller's parameters online.

\section{Considerations and Limitations}

\subsection{Practical Considerations}

Safety is a crucial consideration when applying RL to real-world motor control. As mentioned above, incorporating safety measures is essential to stabilize motor systems during training. Approaches such as safeguarding layers, and reward shaping play a critical role in preventing unsafe actions that could damage the motor or the control system.

Computational complexity is also a key factor, especially for RL algorithms that use deep neural networks. The significant computational cost of these methods requires careful allocation of hardware resources and efficient implementation strategies, particularly in real-time scenarios.

Efficient training methods are also vital for practical application. RL agents generally require extensive data to learn effectively, which may be impractical for real-world motor systems due to time constraints or risks associated with long training duration. 

Lastly, hyperparameter tuning has a substantial impact on RL performance. The effectiveness and stability of RL algorithms are highly sensitive to hyperparameters such as learning rates, exploration noise, and network architecture. Optimizing these parameters often involves considerable experimentation and adjustments to maximize performance.

\subsection{Limitations}

Overfitting to specific scenarios or operating conditions is one of the primary limitations of RL in motor control. RL agents trained on small datasets may fail to generalize and perform in unseen situations. 

Generalizing across different motor types and operating conditions presents another challenge. While some RL algorithms exhibit generalization capabilities, achieving consistent performance across various motor specifications and environmental conditions remains an ongoing challenge.

Finally, RL's dependency on simulation models for training is a limitation in real-world applications. The discrepancies between simulation and real-world systems may result in poor performance when transferring RL agents from the simulation to the actual system. 

\section{Conclusions}

Reinforcement Learning (RL) presents a powerful approach to electric motor control and offers solutions to the limitations of traditional methods, such as PID controllers. RL's model-free characteristic allows it to handle nonlinearities, parameter uncertainties, and external disturbances that usually hinder traditional control methods. The adaptability of RL agents to varying dynamics and their ability to optimize complex control objectives make them specifically useful in complex and nonlinear motor control applications.
However, the computational complexity and hardware requirements associated with RL, especially those using deep neural networks, pose significant challenges. While RL algorithms can achieve outstanding performance, the computational costs and the need for specialized hardware make it difficult for RL to compete with the simplicity and cost-effectiveness of traditional methods, which are often sufficient for less complex applications. RL's adoption will depend on addressing the computational burden and exploring more efficient hardware implementations. Despite all the challenges, RL's potential in motor control remains significant. Further research is necessary to develop more efficient algorithms, optimize training methods, and explore techniques to enhance generalization and real-world applicability. As these efforts progress and hardware costs decrease, RL is going to play an increasingly important role in shaping the future of motor control across different industries.

\bibliography{references}

\end{document}